# Gustav Spörer Was Not a Perfect Observer: Failure of the Active Day Fraction Reconstruction of Sunspot Group Numbers

Leif Svalgaard[1]
Submitted May 2017

**Abstract**

We show that the Active-Day-Fraction calibration method (Willamo et al. [2017]) fails for Gustav Spörer's sunspot group observations. Spörer was labeled a 'perfect observer' on account of his 'observational threshold $S_S$ area' being determined to be equal to zero, based on the assumption that the observer can see and report all the groups with the area larger than $S_S$, while missing all smaller groups. So, Spörer could apparently, according to the ADF calibration method, see and report all groups, regardless of size and should never miss any. This suggests a very direct test: compute the yearly average group count for both Spörer and the 'perfect observer' exemplar, the Royal Greenwich Observatory (RGO), and compare them. They should be identical within a reasonable (very small) error margin. We find that they are not and that RGO generally reported 45% more groups than Spörer, and that therefore, the ADF-method is not generally applicable.

## 1. Introduction

In a recent article, Willamo et al. [2017] upgrade the Sunspot Group Number reconstruction based on the fraction of 'Active Days' per month suggested by Vaquero et al. [2012] and extended by Usoskin et al. [2016] and touted as a "modern non-parametric method […] free from daisy-chaining and arbitrary choices". If this were indeed the case, significant progress would have been made in the quest for the elusive record of long-term solar activity. We show in this brief communication that the method fails spectacularly when applied to what the article calls "a perfect observer", and that therefore the jury is still out on this, rendering the vaunted ADF-methodology suspect and less than useful.

## 2. The Data

We concentrate on the interval 1880-1893 where sufficient and unambiguous data are available from the following observers: Gustav Spörer (at Anclam), Royal Greenwich Observatory (RGO), and Alfred Wolfer (Zürich), as provided by Usoskin (Personal Communication, 2017 to Laure Lefèvre) in this format:

```
Year M D  G  G(ADF)  GLo GHi
1880 1 4  1  1.04806  1   1
1880 1 7  2  2.07032  2   2
1880 1 8  3  3.09613  3   3
```

Year, M=Month, D=Day, G=Observed group count
G(ADF)= ADF-based reconstruction
$G_{Lo}$=Low Limit of G(ADF)
$G_{Hi}$=High Limit of G(ADF)

---

[1] Stanford University, Stanford, CA 94305, USA

It is not clear from the data if the limits $G_{Lo}$ and $G_{Hi}$ (determining the confidence interval) are truncated or rounded to the nearest integer or if they are the actual true values. In any case, they are always identical for Spörer.

Table 1 of Willamo et al. [2017] specifies that Spörer is a 'perfect observer' with 'observational threshold $S_S$ (in millionths of the solar disk)' equal to zero, based on the assumption that the 'quality' of each observer is characterized by his/her observational acuity, measured by a threshold area $S_S$. The threshold implies that the observer can see and report all the groups with the area larger than $S_S$, while missing all smaller groups. So, Spörer could apparently, according to the ADF calibration method, see and report all groups, regardless of size and should never miss any, except for a few that evolved and died without Spörer seeing them. In fact, the $G_{Lo}$ and $G_{Hi}$ given by Usoskin are identical as they should be for perfect data without errors. If so, it suggests a very direct test: compute the yearly average group count for both Spörer and RGO and compare them. They should be identical within a reasonable (very small) error margin.

## 3. The Test

The following table gives the annual values for Spörer (calculated by Willamo et al. [2017]), Spörer (observed and reported), RGO, Wolfer, and the Svalgaard & Schatten [2016] Group Number Backbone:

| Year | Spörer(W) | Spörer(O) | RGO | Wolfer | S&S BB |
|------|-----------|-----------|------|--------|--------|
| 1880.5 | *2.18* | 2.11 | 2.19 | 2.69 | 2.70 |
| 1881.5 | 3.11 | 3.03 | 3.96 | 4.69 | 4.62 |
| 1882.5 | 3.56 | 3.46 | 4.48 | 4.59 | 4.78 |
| 1883.5 | 3.57 | 3.47 | 4.92 | 5.90 | 5.31 |
| 1884.5 | 3.87 | 3.78 | 5.58 | 5.53 | 5.84 |
| 1885.5 | 2.89 | 2.81 | 4.28 | 4.32 | 4.64 |
| 1886.5 | 1.93 | 1.87 | 2.04 | 2.17 | 2.41 |
| 1887.5 | 1.17 | 1.12 | 1.25 | 1.44 | 1.35 |
| 1888.5 | 0.61 | 0.57 | 0.72 | 0.73 | 0.78 |
| 1889.5 | 0.32 | 0.29 | 0.52 | 0.60 | 0.60 |
| 1890.5 | 0.59 | 0.55 | 0.71 | 1.15 | 0.69 |
| 1891.5 | 2.58 | 2.51 | 3.41 | 4.17 | 3.56 |
| 1892.5 | 4.08 | 3.98 | 6.39 | 5.98 | 6.18 |
| 1893.5 | 5.62 | 5.50 | 8.51 | 8.31 | 7.73 |
| Average | 2.577 | 2.504 | 3.497 | 3.733 | 3.656 |
| Ratio | 1.029 | 1.000 | 1.397 | 1.491 | 1.460 |

We here posit that what Spörer actually reported (column three) is what must be compared to the reconstructions. It is thus evident that RGO is 40%, Wolfer 49%, and S&S BB 46% higher than what Spörer 'the perfect observer' saw and reported. And that therefore the test has failed. The ADF-method of calibration does not give the correct result in this simple, straightforward, and transparent example.

Often, a picture is worth a thousand words, so Figure 1 shows the results in graphical form.

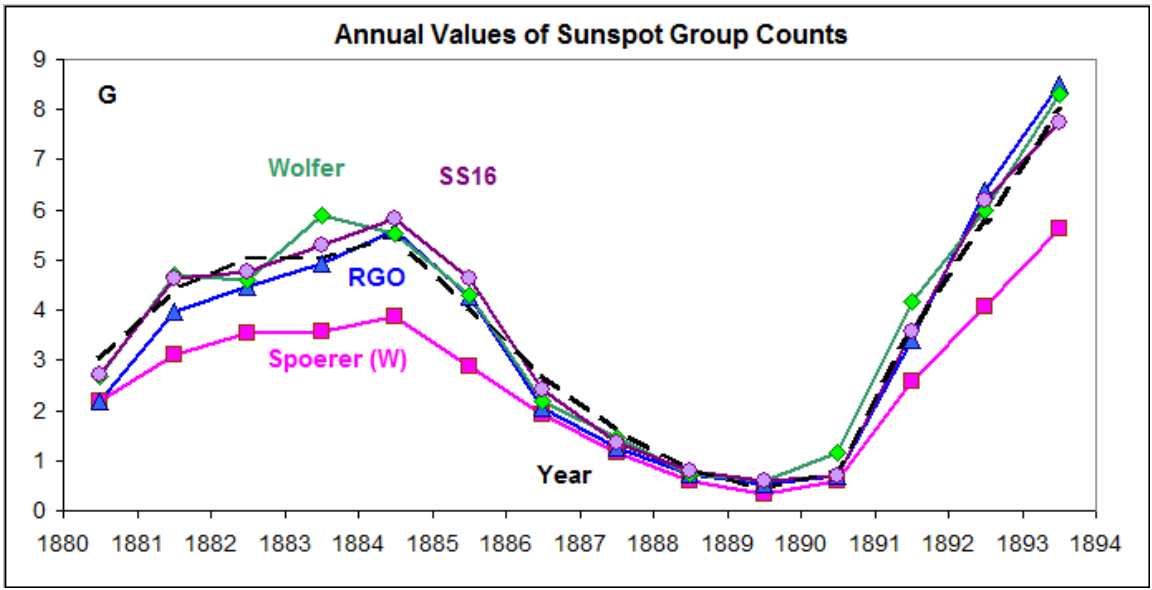

**Figure 1.** Annual values of the Sunspot Group Number for Spörer (pink squares; calculated by Willamo et al. [2017]), RGO (blue triangles), Wolfer (green diamonds), Svalgaard & Schatten [2016] (purple dots). Scaling Spörer up by a factor 1.45 yields the black dashed curve.

The difference between Spörer and the real 'perfect observer' RGO is vividly evident in Figure 2 that shows the fraction of the time where a given number of groups was observed as a function of the phase within the sunspot cycle. At high solar activity Spörer saw significantly fewer spots than RGO. It is also at such times that the ADF is close to unity (as at such times almost every day is an 'active day' in every cycle) and therefore does not carry information about the size of the cycle.

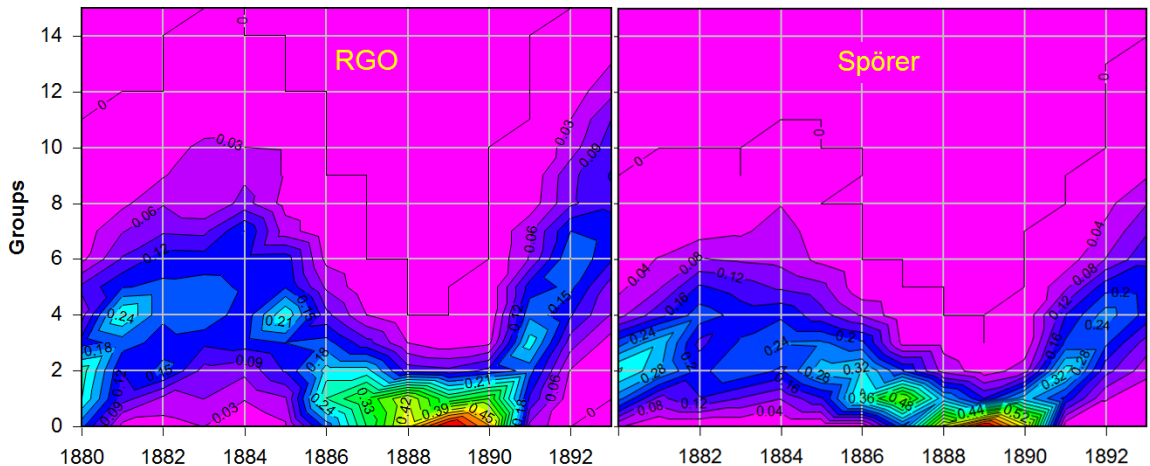

**Figure 2.** Frequency of occurrence of counts of groups on the solar disk as a function of time during 1880-1893 for RGO (left) and Spörer (right) determined for each year by the number of days where a given number of groups was observed on the disk divided by the number of days with an observation.

## 4. Conclusion

The ADF-method does not yield a correct 'observational threshold $S_S$' for G. Spörer and thus does not form a reliable basis for reconstruction of past solar activity valid for all times and observers, and as such must be discarded for general use.

## Acknowledgements

I thank Ed Cliver to drawing attention to the designation of Spörer as a 'perfect observer'. I thank Stanford University for continuing support. I also thank Laure Lefèvre for providing Ilya Usoskin's data files pertaining to the ADF method.